\documentclass[twocolumn,nofootinbib,aps,amsmath,superscriptaddress,preprintnumbers]{revtex4}

\usepackage{epsfig}
\usepackage{graphicx}
\usepackage[hypertex]{hyperref}

\newcommand{\nn}{\nonumber}
\newcommand{\plus}{\ensuremath{\! + \!}}
\newcommand{\minus}{\ensuremath{\! - \!}}

\def\OMIT#1{}

\newcommand{\OLam}{{\cal O}(\Lambda_{\rm QCD})}

\begin{document}

\preprint{\vbox{\hbox{MIT--CTP 3940}  \hbox{MPP-2008-24} 
\hbox{arXiv:0803.4214}
}}

\title{\boldmath Infrared Renormalization Group Flow for Heavy Quark Masses 
\vspace{0.1cm}}

\author{Andr\'e H. Hoang}
  \affiliation{Max-Planck-Institut f\"ur Physik (Werner-Heisenberg-Institut) 
  F\"ohringer Ring 6, 80805 M\"unchen, Germany\vspace{0.0cm}}

\author{Ambar Jain}
\affiliation{Center for Theoretical Physics, Massachusetts Institute of
Technology, Cambridge, MA 02139\vspace{0.0cm}}

\author{Ignazio Scimemi}
\affiliation{Departamento de Fisica Teorica II,
Universidad Complutense de Madrid,
28040 Madrid, Spain\vspace{0.2cm}}

\author{Iain W.\ Stewart\vspace{0.2cm}}
\affiliation{Center for Theoretical Physics, Massachusetts Institute of
Technology, Cambridge, MA 02139\vspace{0.0cm}}

\begin{abstract}
  
  A short-distance heavy quark mass depends on two parameters, the
  renormalization scale $\mu$ controlling the absorption of ultraviolet
  fluctuations into the mass, and a scale $R$ controlling the absorption of
  infrared fluctuations. $1/R$ can be thought of as the radius for perturbative
  corrections that build up the mass beyond its point-like definition in the
  pole scheme. Treating $R$ as a variable gives a renormalization group
  equation. We argue that the sign of this anomalous dimension is universal:
  increasing $R$ to add IR modes decreases $m(R)$.  The flow improves the
  stability of conversions between mass schemes, allowing us to avoid large logs
  and the renormalon. The flow in $R$ can be used to study IR renormalons
  without using bubble chains, and we use it to determine the coefficient of the
  ${\cal O}(\Lambda_{\rm QCD})$ renormalon ambiguity of the pole mass with a
  convergent sum-rule.
 
\end{abstract}

\maketitle

The pole-mass, $m_{\rm pole}$, provides a simple definition of a mass-parameter
in perturbative quantum field theory, corresponding to the location of the
single particle pole in the two-point function. For the electron mass in QED
$m_{\rm pole}$ is used almost exclusively, but for quarks in QCD there are two
reasons it is impractical.  First, at high energies, large logs appear which
spoil perturbation theory with $m_{\rm pole}$. This problem is cured by
introducing the concept of a running-mass $m(\mu)$, where the renormalizaton
group (RG) flow in $\mu$ is controlled by a mass-anomalous dimension. The
second, and more serious problem, is that due to confinement there is no pole in
the quark-propagator in non-perturbative QCD.  Thus the concept of a quark
pole-mass is ambiguous by $\Delta m_{\rm pole}\sim \Lambda_{\rm QCD}$. This
ambiguity appears as a linear sensitivity to infrared momenta in Feynman
diagrams, and results in a diverging perturbation series for any observable
expressed in terms of $m_{\rm pole}$, with terms $\sim 2^n n!\,\alpha_s^{n+1}$
asymptotically for large $n$. For the heavy quark masses (charm, bottom, top)
that we study, this behavior is referred to as the pole-mass ${\cal
  O}(\Lambda_{\rm QCD})$ renormalon problem~\cite{Bigi:1994em}, where the Borel
transform of the series has a singularity at $u=1/2$.  Schemes without this
infrared problem are known as short-distance masses, and always depend on an
additional infrared scale $R$.

Typically, $R$ is considered as intrinsic to the short-distance quark mass
definition, $m_R(\mu)$.  Examples are
\begin{align} \label{masses}
  {\overline {\rm MS}} &:  && {\overline m}(\mu),    & R &= {\overline m}(\mu) 
   ; \\
  {\rm RGI}~[2] &: && m_{\rm RGI}, & R &= m_{\rm RGI} 
   ; \nn \\
  {\rm kinetic}~[3] &: && m_{\rm kin}, & R & = \mu_f^{\rm kin} ;
  \nn\\
  {\rm 1S}~[4] &:   && m_{\rm 1S},   & R & = m_{\rm 1S} C_F \alpha_s(\mu) ; 
  \nn\\
  {\rm PS}~[5] &:   && m_{\rm PS},   & R & = \mu_f^{\rm PS} \,. \nn
\end{align}
where $C_F=4/3$.  Many schemes have $R=m$, but this is not generic. For
instance, the 1S-mass is defined as one-half the mass of the heavy quarkonium
${}^3S_1$ state in perturbation theory, and its $R$ is of order the inverse Bohr
radius. In the kinetic and the potential subtraction (PS) schemes $R$ is set by
cutoffs, $\mu_f^{\rm kin}$ and $\mu_f^{\rm PS}$, on integrals over a heavy-quark
correlator and the heavy-quark static potential respectively.  Depending on the
scales involved in a process, schemes with a specific range of $\mu$ and $R$ are
most appropriate to achieve stable perturbative results.

The goal of this letter is to consider $R$ as a continuous parameter, and study
the RG flow in $R$ of masses $m(R,\mu)=m_R(\mu)$. We consider converting between
mass schemes $m_A(R,\mu)$ and $m_B(R',\mu)$ where $R \ll R'$. To avoid the
$\OLam$ renormalon in fixed-order perturbation theory a common expansion in
$\alpha_s(\mu)$ must be used, which inevitably introduces large logs,
$\ln(R'/R)$. The RGE in $R$ allows mass-scheme conversions to be done avoiding
both large logs and the renormalon. We show this improves the stability of
conversions between the $\overline {\rm MS}$ scheme with $R=m$, and low energy
schemes with $R \ll m$ that are extensively used for high precision
determinations of heavy quark masses~\cite{PDG}.  The solution of this RGE is
also used to systematically derive a convergent series for the normalization of
the $u=1/2$ singularity in the pole-mass Borel transform.

To start, translate the bare-quark mass in QCD to the pole-mass, $m_{\rm bare} =
Z_m m_{\rm pole}$, where UV divergences from scales $p^2\gg m^2$ appear in the
mass-renormalization constant $Z_m$. The difference between using $m_{\rm pole}$
and any other scheme $m(R,\mu)$ corresponds to specifying additional finite
subtractions, $\delta m(R,\mu)$. Let
\begin{align} \label{dm}
 m_{\rm pole} &= m(R,\mu) + \delta m(R,\mu) \,,\\
 \delta m(R,\mu) & = R  \sum_{n=1}^\infty \sum_{k=0}^n  
  a_{nk}\: \Big[\frac{\alpha_s(\mu)}{4\pi}\Big]^n 
  \ln^k\!\Big(\frac{\mu}{R}\Big) 
 \,. \nn
\end{align}
Here $a_{nk}$ are numbers, and $\alpha_s$ is in the $\overline {\rm
  MS}$-scheme with
\begin{align} \label{muAD}
 \frac{d\alpha_s(\mu)}{d\ln\mu} &= \beta[\alpha_s(\mu)] 
   = -2 \alpha_s(\mu) \sum_{n=0}^\infty \beta_n \,
   \Big[\frac{\alpha_s(\mu)}{4\pi}\Big]^{n+1}
  \,.
\end{align}
We will only consider gauge independent short-distance mass schemes for
$m(R,\mu)$, where $\delta m$ eliminates the infrared ambiguity associated to
the pole mass. This requires that $a_{(n+1)0}\sim 2^{n} n! $ asymptotically for
large $n$.

These mass schemes come in two categories. In $\mu$-independent schemes (such as
RGI, kinetic, 1S):
$\frac{d }{d\ln \mu} \: m(R,\mu) = 0$.
In these schemes $a_{11}=0$, and $a_{nk}$ with $k\ge 1$ are determined by the
$a_n\equiv a_{n0}$'s and $\beta_n$'s. Thus the $a_{n}$'s specify the scheme.
Masses in the other category have a $\mu$-anomalous dimension (like $\overline
{\rm MS}$), and using $d/d\ln\mu\: m_{\rm pole}=0$ one finds
$ \frac{d}{d\ln\mu}\: m(R,\mu) = 
 - R\: \gamma_\mu[\alpha_s(\mu)]$.
 Here $a_{n1}$ and $a_{n0}$ are needed to specify the scheme and $\gamma_\mu$.
 $\gamma_\mu$ does not depend on $\ln(\mu/R)$, so all $a_{nk}$ with $k\ge 2$ are
 determined. For ``mass-independent'' schemes like $\overline {\rm MS}$ we
 always have $a_{11}= 6 C_F$, and a universal $\gamma_\mu$ at leading order
 (LO).

Eq.~(\ref{dm}) can be used to identify $R$ for schemes like those in
Eq.~(\ref{masses}). To see that $R$ is related to absorbing IR fluctuations into the
mass, consider the PS scheme where
\begin{equation}
  m^{\rm PS}(R) -m_{\rm pole} = -\delta m^{\rm PS}(R) \equiv
  \frac{1}{2}\,
  \raisebox{-0.4cm}{$ \stackrel{\mbox{\LARGE $\int$}}{\mbox{\scriptsize $|q|\!\!<\!\!R$}} $}
  \:
  \frac{d^3q}{(2\pi)^3}\, V(q) \,.
\end{equation}
Here $V(q)$ is the momentum space color singlet static potential between
infinitely heavy test charges in the ${\bf 3}$ and $\bar {\bf 3}$
representations. In $m_{\rm pole}-\delta m^{\rm PS}$ the low momentum part of
the potential precisely cancels the infrared sensitivity of $m_{\rm pole}$,
leaving a well-defined short-distance mass, $m^{\rm PS}$. If
we increase $R$ from $R_0$ to $R_1$ then
\begin{align} \label{R0R1}
-\delta m^{\rm PS}(R_1) &= 
\int_0^{R_0} \!\!\!\! {dq}\ \frac{q^2\:V(q)}{4\pi^2} 
 \: + \!
 \int_{R_0}^{R_1} \!\!\!\! {dq}\ \frac{q^2\:V(q)}{4\pi^2} 
\,,
\end{align}
so additional potential energy is absorbed into $m^{\rm PS}$, increasing the
range of IR fluctuations included in the PS-mass.  In other mass-schemes the
precise definition of $R$ differs, but the interpretation of this scale as an
IR-cutoff still remains. Another simple example is what we call the
static-scheme.  The static energy, $E^{\rm static}(r)= 2 m_{\rm pole} + V(r)$ is
free of the ${\cal O}(\Lambda_{\rm QCD})$ renormalon, which requires a
cancellation of IR sensitivity between the pole-mass and the potential energy
$V(r)$. Here $V(r)$ is the Fourier transform of $V(q)$. A short-distance static
mass, $m^{\rm stat}(R)$, can be defined to make this cancellation explicit, with
$ \delta m^{\rm stat}(R) = -\frac12 \: V(r=1/R)$.  [We use the modern definition
of the static potential, which becomes $\mu$-dependent at ${\cal
  O}(\alpha_s^4)$~\cite{Brambilla:1999qa}, but does not suffer from infrared
divergences.  Since ${\cal O}(\alpha_s^4)$ is beyond the order of our analysis
we refer to the PS and static masses as $\mu$-independent.  For convenience we
also consider the kinetic scheme in the heavy-quark limit, so there are no
higher powers of $R$ in Eq.~(\ref{dm}).] Since $V$ is attractive, increasing $R$
decreases the PS and the static mass. We will see that this decrease is
universal. It is described by an RGE in $R$ where $d/d\ln\! R\: m(R)<0$ at LO in
$\alpha_s(R)$ for all known physical mass schemes.

{\bf The RGE for $\mathbf{R}$}.  Consider any $\mu$-independent scheme where $R$
is a free parameter, such as the PS, static, and kinetic schemes.  The pole-mass
in Eq.~(\ref{dm}) is $R$-independent, so $R d/dR\, m(R) = - R d/dR\, \delta
m(R)$.  To avoid having large $\ln(\mu/R)$'s on the RHS we must expand in
$\alpha_s(R)$, $\delta m(R) = R \sum_{n=1}^\infty a_{n}\:
\big[{\alpha_s(R)}/{(4\pi)} \big]^n$.  This yields an RGE for $R$
\begin{align} \label{RRGE}
  R \frac{d}{dR} m(R) &= - \frac{d}{d\ln R} \delta m(R) 
  \equiv - R \: \gamma_R[\alpha_s(R)] \,,
   \nn\\
  \gamma_R[\alpha_s(R)] &= \sum_{n=0}^\infty \gamma^R_n \: 
  \Big[ \frac{\alpha_s(R)}{4\pi}\Big]^{n+1}
  \,.
\end{align}
For the kinetic scheme the RGE in $R$ was formulated in
Refs.~\cite{Voloshin:1992wg,Bigi:1997fj}. To our knowledge the full implications
of Eq.~(\ref{RRGE}) have not yet been studied. We will refer to $\gamma_0^R$,
$\gamma_1^R$, $\gamma_2^R$ as the LO, NLO, NNLO, anomalous dimensions of the RGE
in $R$. Here $\gamma_0^R = a_1$, $\gamma_1^R = a_2 -2 \beta_0 a_1$, $\gamma_2^R
= a_3 -4 \beta_0 a_2 -2\beta_1 a_1$, so they are determined by the
non-logarithmic terms in Eq.~(\ref{dm}). 
For $\mu$-dependent renormalization schemes with a parameter $R$, the same RGE
in Eq.~(\ref{RRGE}) is obtained once we set $\mu=R$ and define $m(R)=m(R,R)$ and
$\delta m(R)=\delta m(R,R)$.  These schemes have consistent RG flow in the
two-dimensional $R$--$\mu$ plane (vertically and along the diagonal).  An
example of such a scheme is the jet-mass, $m^{\rm jet}(R,\mu)$, defined via the
position-space jet-function~\cite{Jain:2008gb}. To estimate uncertainties in the
RGE in $R$ one can set $\mu=\kappa R$ in $\delta m$ and determine
$\gamma^R_n(\kappa)$, then vary about $\kappa=1$.

Since the $\OLam$ ambiguity in $m_{\rm pole}$ is $R$-indepen\-dent, the
derivative of $\delta m$ in Eq.~(\ref{RRGE}) ensures that $\gamma_R[\alpha_s]$
does not contain this renormalon (a key point!).  The RGE flow in the parameter
$R$ takes us from a renormalon free mass $m(R_0)$ to the renormalon free mass
$m(R_1)$.

Examples of LO anomalous dimensions are
\begin{align}
 & (\gamma_0^R)^{\rm PS} = 4C_F \,,
  \quad (\gamma_0^R)^{\rm stat} = 2\pi C_F \,,
  \quad  (\gamma_0^R)^{\rm MSR} = 4 C_F  \,,  \nn\\
 & 
  (\gamma_0^R)^{\rm kinetic} = 16 C_F/3 \,,
  \quad (\gamma_0^R)^{\rm jet} = 2 e^{\gamma_E} C_F \,.
\end{align}
One can find another suitable scheme from the $\overline {\rm MS}$--pole mass
relation, by taking $\overline m(\overline m)\to R$ in $\delta m$. We call this
the MSR-scheme.  All these $\gamma_0^R$'s are positive. Thus for large enough
$R$ we have $d/d\ln\! R\: m(R)<0$, and increasing $R$ always decreases $m(R)$.
This sign appears as a universal feature of physical short-distance mass
schemes. Now, the norm of $\gamma_0^R$ does depend on the scheme. For a given
change $\Delta R$ it determines the amount of IR fluctuations that are added to
$m(R)$.  In a different scheme an equivalent amount of IR fluctuations can
always be added to the mass with a different change $\Delta R'$. To see this,
consider rescaling $R= \lambda R'$ with a $\lambda>0$. We demand $\lambda \sim
{\cal O}(1)$ to avoid large logs.  Expanding $\alpha_s(\lambda R') =
\alpha_s(R')- \beta_0 \ln\lambda\, \alpha_s^2(R')/(2\pi) +\ldots$ gives
\begin{align}\label{lambda}
  \gamma_0^{R'} &= \lambda \gamma_0^R \,, 
  \ \
  \gamma_1^{R'} = \lambda \big[\gamma_1^R - 2\beta_0 \gamma_0^R \ln\lambda
  \big]  \,, \\
  \gamma_2^{R'} &= \lambda \big[ \gamma_2^R
    - (4\beta_0 \gamma_1^R+2\beta_1\gamma_0^R) \ln\lambda 
    + 4 \beta_0^2 \gamma_0^R \ln^2\lambda
    \big] \,.\nn
\end{align}
Thus at LO a scale change in $R$ just modifies the norm of $\gamma_0^R$, and we
are free to pick $\lambda$ so that $\gamma_0^{R'}$ is equal to the LO anomalous
dimension in some other scheme. The condition $(\gamma_0^{R'}/\gamma_0^R)\sim
{\cal O}(1)$ identifies a class of related schemes parameterized by the scale
choice for $R$ (our $\lambda$).
For a top-quark in the PS, static, and MSR schemes we find $\{\tilde
\gamma_0^R$, $\tilde \gamma_1^R$, $\tilde \gamma_2^R\}=$
$\{0.348$,$0.108$,$0.231\}^{\rm PS}$, $\{0.546$, $-0.061$,$0.143\}^{\rm stat}$,
$\{0.348$,$0.213$,$0.068\}^{\rm MSR}$, where we let $\tilde \gamma_k^R\equiv
\gamma_k^R/(2\beta_0)^{k+1}$ and used
Refs.~\cite{Peter:1996ig,Melnikov:2000qh,PDG}.  Here and below we use 5 light
running flavors.


To determine the general solution to Eq.~(\ref{RRGE}) first write
\begin{align}
 \ln\frac{R_1}{R_0} =\!\! \int_{\alpha_0}^{\alpha_1}\!\! \frac{d\alpha_R}{\beta[\alpha_R]} 
  =\! \int_{t_1}^{t_0}\!\!\! dt\: \hat b(t) = G(t_0) - G(t_1) \,,
\end{align}
where $\alpha_i\equiv \alpha_s(R_i)$, $\alpha_R\equiv \alpha_s(R)$, $t_i \equiv
-2\pi/(\beta_0\alpha_i)$, and $t\equiv -2\pi/(\beta_0\alpha_R)$. For the first
few orders
\begin{align} \label{baexpn}
  \hat b(t) &= 1 +\frac{\hat b_1}{t} + \frac{\hat b_2}{t^2} +
  \frac{\hat b_3}{t^3} \ldots \,,\nn\\
   G(t) &= t + \hat b_1 \ln(-t) - \frac{\hat b_2}{t}- \frac{\hat b_3}{2t^2}
   - \ldots \,, 
\end{align}
where $\hat b_1 = \beta_1/(2\beta_0^2)$, $\hat
b_2=(\beta_1^2-\beta_0\beta_2)/(4\beta_0^4)$, and $\hat
b_3=(\beta_1^3-2\beta_0\beta_1\beta_2+\beta_0^2\beta_3)/(8\beta_0^6)$. Note that
$G'(t) =\hat b(t)$.  The function $G(t)$ allows us to define
\begin{align} \label{LQCD}
 \Lambda_{\rm QCD} = R\, e^{G(t)} = R_0\, e^{G(t_0)} \,,
\end{align}
a definition that is valid to any order in perturbation theory, and corresponds
to the familiar definition of $\Lambda^{(k)}_{\rm QCD}$ at ${\rm N}^k{\rm LL}$
order.  Now by making a change of variable we obtain the all orders result
\begin{align} \label{Rsoln}
 m(R_1)-m(R_0) &= -\int_{\ln R_0}^{\ln R_1} \!\!\!\! d\ln\! R
  \ \: R\, \gamma_R[\alpha_s(R)] \\
  &= \Lambda_{\rm QCD} \int_{t_1}^{t_0} \!\!\! dt\:
   \gamma_R(t)\: \frac{d}{dt}\: e^{-G(t)} \,, \nn
\end{align}
where $\gamma_R(t)= \gamma_R[\alpha_s(R)]$.  This is a very convenient formula
for the solution of the RGE in $R$.

Lets consider the LL solution. The RGE is
\begin{align}
  R \frac{d}{dR} m(R) = -  R\, \gamma_0^R\: \frac{\alpha_s(R)}{4\pi}  \,.
\end{align}
Here $\gamma_R(t)= -\gamma_0^R/(2\beta_0 t)$ and with the leading terms for $\hat
b(t)$ and $G(t)$, Eq.~(\ref{Rsoln}) gives
\begin{align} \label{LLsoln}
  m(R_1)-m(R_0) &
  =\frac{\Lambda_{\rm QCD}^{(0)}\,\gamma_0^R}{2\beta_0}
   \int_{t_1}^{t_0} \!\! dt\: \frac{e^{-t}}{t} 
 \\
  &=
  \frac{\Lambda_{\rm QCD}^{(0)}\,\gamma_0^R}{2\beta_0}
   \big[ \Gamma(0,t_1) - \Gamma(0,t_0) \big]
  \nn \,.   
\end{align}
Since $t_1 < t_0 < 0$ the integral is convergent.  Here and below we make use of
the incomplete gamma function
\begin{align}
  \Gamma[c,t] = \int_t^\infty \!\! dx \ \: x^{c-1}\: e^{-x} \,,
\end{align}
and note that differences like the one in Eq.~(\ref{LLsoln}) have no
contributions from the cut present in $\Gamma[c,t]$ for $t<0$.  To see which
perturbative terms the solution in Eq.~(\ref{LLsoln}) contains, recall the
asymptotic expansion for $t\to \infty$
\begin{align} \label{Gammact}
 \Gamma[c,t] \stackrel{\rm asym}{=} 
  e^{-t}\, t^{c-1} \sum_{n=0}^\infty 
  \frac{\Gamma(1\minus c\plus n)}{(-t)^n\, \Gamma(1\minus c)}
  \,.
\end{align}
For $c=0$ this plus the LL relation $\Lambda^{(0)}_{\rm QCD}e^{-t}=R$ yields
\begin{align}
 \Lambda_{\rm QCD}^{(0)} \Gamma[0,t]  & \stackrel{\rm asym}{=} 
  -2R \sum_{n=0}^\infty  2^n\, n!\, \Big[\frac{\beta_0\alpha_s(R)}{4\pi}\Big]^{n+1} \,.
\end{align}
This is a divergent series, but for Eq.~(\ref{LLsoln}) we have 
\begin{align}\label{expnmm}
  &m(R_1)-m(R_0) \\
  &= \frac{-\gamma_0^R R_1 }{2\beta_0} \sum_{n=0}^\infty 
   \Big[\frac{\beta_0\alpha_1}{2\pi}\Big]^{n+1}  n!\, \bigg( 1 \!-\! \frac{R_0}{R_1} 
   \sum_{k=0}^n \frac{1}{k!} \ln^k\!\frac{R_1}{R_0}\bigg) 
  \nn \\
  & = -\frac{\gamma_0^R R_0 }{2\beta_0}  \sum_{n=0}^\infty 
   \Big[\frac{\beta_0\alpha_1}{2\pi}\Big]^{n+1}  
   \sum_{k=n+1}^\infty \frac{n!}{k!} \ln^k\!\frac{R_1}{R_0} \,,\nn
\end{align}
which is convergent since $\beta_0\alpha_s(R_1)\ln(R_1/R_0)/(2\pi) <1$.
Eq.~(\ref{expnmm}) displays the problem of large logs in fixed order
perturbation theory for $R_1\gg R_0$.  The RGE in $R$ encodes IR physics from
the large order behavior of perturbation theory.  It causes a rearrangement of
the IR fluctuations included in the mass in going from $R_0$ to $R_1$ without
reintroducing a renormalon.

Using Eq.~(\ref{Rsoln}) the solution in Eq.~(\ref{LLsoln}) can be extended to
include all the terms up to ${\rm N}^k{\rm LL}$. Let $[t \gamma_R(t)\, \hat b(t)\,
e^{-G(t)}\, e^t\, (-t)^{\hat b_1}] \equiv -\sum_{j=0}^\infty S_j \:
(-t)^{-j}$, then 
\begin{align} \label{NkLLsoln}
 & \big[m(R_1)\minus m(R_0)\big]^{{\rm N}^k{\rm LL}} 
  = \Lambda^{(k)}_{\rm QCD}\ \mbox{$\sum_{j=0}^k$}\  S_j\, (-1)^j \nn\\
  & \qquad
  \times e^{i\pi \hat b_1}  \big[ \Gamma(-\hat b_1\minus j,t_1) 
  - \Gamma(-\hat b_1 \minus j, t_0) \big]
  \,.
\end{align}
This solution is real.  Here $\Lambda_{\rm QCD} \Gamma[c,t] \sim R\, {\cal
  O}(\alpha_s^{1-c})$ encodes the suppression of higher order terms, see
Eq.(\ref{Gammact}).  Using $\tilde \gamma_k^R\equiv
\gamma_k^R/(2\beta_0)^{k+1}$ the first few coefficients are
\begin{align}
 S_0 &= \tilde \gamma_0^R \,, 
  \quad
 S_1 = \tilde \gamma_1^R - (\hat b_1\plus \hat b_2) \tilde \gamma_0^R\,,
  \\
 S_2 &= \tilde \gamma_2^R - (\hat b_1\plus \hat b_2) \tilde \gamma_1^R
    +\big[ (1\plus \hat b_1)\hat b_2 +(\hat b_2^2 \plus \hat b_3)/2 \big]
  \tilde \gamma_0^R .
  \nn
\end{align}
%

{\bf Connection to $\mathbf m_{\rm pole}$}.  Eq.~(\ref{Rsoln}) allows us to
study the $u=1/2$ renormalon in $m_{\rm pole}$. This is done by taking the limit
$R_0\to 0$ and $t_0= -\ln(R_0/\Lambda_{\rm QCD})+\ldots \to\infty$, where we
continue in the upper complex plane around the Landau pole.  Since $\lim_{R_0\to
  0} m(R_0) = m_{\rm pole}$, we obtain
\begin{align} \label{mRtompole}
  m(R_1)- m_{\rm pole} &= \Lambda_{\rm QCD} \int_{t_1}^\infty \!\!\!\! dt\ 
    \gamma_R(t)\:  \frac{d}{dt}\, e^{-B(t)} \,.
\end{align}
Upon expansion in $\alpha_s$ the RHS reproduces $(-\delta m)$ in Eq.~(\ref{dm}).
Eq.~(\ref{mRtompole}) represents an all order expression for the $\Lambda_{\rm
  QCD}$ renormalon in $m_{\rm pole}$. This result can be manipulated into an
equivalent expression for an inverse Borel transform of $B(u)$. We obtain
%
\begin{align} \label{Bu}
 & m(R) -m_{\rm pole} = \int_0^\infty \!\! du \ B(u)\ e^{-u
    \frac{4\pi}{\beta_0\alpha_s(R)}} \,, \\
& B(u) = 2 R\bigg[ \sum_{\ell=0}^\infty g_\ell\, Q_\ell(u) 
  - P_{1/2} \sum_{\ell=0}^\infty g_\ell\,\frac{ \Gamma(1\plus \hat
 b_1 \minus \ell)}{(1\minus 2u)^{1+ \hat b_1-\ell}}
  \bigg],
 \nn \\ \label{Bup}
 & P_{1/2}  = \sum_{k=0}^\infty   \ 
\frac{S_k }{\Gamma(1\plus \hat b_1\plus k)}\,, 
\end{align}
\begin{align} \label{Bup} 
& Q_\ell(u) = \sum_{k=0}^\infty \frac{S_k (2u)^{k+\ell}
   {}_2F_1(1,1\plus \hat b_1\plus k,2\plus \hat b_1\minus \ell,1\minus 2u)}
   {(1\plus \hat b_1 \minus \ell)\,\Gamma(k+\ell)}
  .
  \nn
\end{align}
Here
\begin{figure}[t!]
\centerline{\includegraphics[width=0.8\columnwidth]{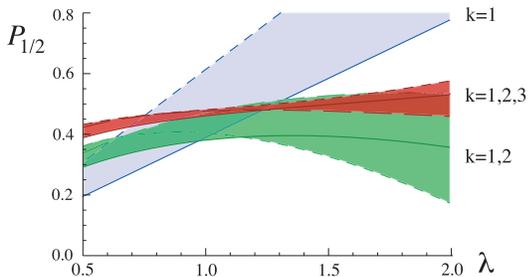}}
\vskip -0.2cm
\caption{Convergence of the sum-rule for $P_{1/2}$ for $m_{\rm pole}$ .}
\label{sumrule}
\vskip-0.4cm
\end{figure}
$ e^{G(t)}\: e^{-t}\: (-t)^{-\hat b_1} \equiv \sum_{\ell=0}^\infty g_\ell
\: (-t)^{-\ell}$, so $g_0=1$, $g_1=\hat b_2$, $g_2=(\hat b_2^2-\hat b_3)/2$,
etc. The normalization $P_{1/2}$ multiplies all terms singular at $u=1/2$ in
Eq.~(\ref{Bu}).  Since $\gamma_R(t)$ is free of the $u=1/2$ renormalon, the
large order behavior of $\gamma_k^R$ is dominated by the next pole at $u=\rho
> 1/2$. This implies that asymptotically for large $k$, $\gamma_k^R\sim k!\,
(\beta_0)^{k+1} \rho^{-k}$. Given that the sum for $\beta[\alpha_s]$ (and hence
$\sum_{\ell} g_\ell$) converges, $S_k \sim \tilde \gamma_k\sim k!\, (2
\rho)^{-k}$, so the sum over $k$ in $P_{1/2}$ converges.  Since
$Q_\ell(1/2)=\sum_{k=0}^\infty S_k [(1\plus \hat b_1\minus \ell)\Gamma(k\plus
\ell)]^{-1}$, all sums over $k$ are {\it absolute convergent} for $u$ close to
$1/2$.  From Eq.~(\ref{Bu}) the large-$n$ asymptotic behavior for any
$m(R)-m_{\rm pole}$ is
$a_{n+1} \sim a_{n+1}^{\rm asym}\equiv P_{1/2}\,  (2\beta_0)^{n+1}\,
\mbox{$\sum_{\ell=0}^\infty$}\: g_\ell\, \Gamma(1\plus \hat b_1\minus \ell+n),$
%
\begin{figure}[t!]
\centerline{\includegraphics[width=0.8\columnwidth]{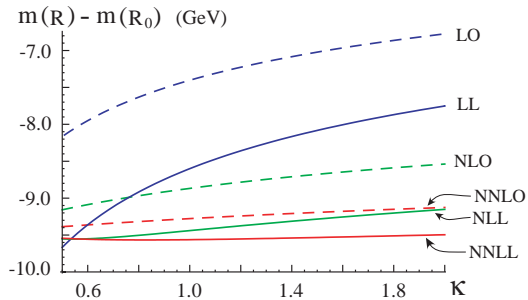}}
\vskip -0.2cm
\caption{Top-mass scheme conversion from $R_0=3\,{\rm GeV}$ to $R=163\,{\rm
    GeV}$. Shown are fixed order results (LO,NLO,NNLO) and RGE results
  (LL,NLL,NNLL), both in the MSR scheme. }
\label{rgefixed}
\vskip-0.4cm
\end{figure}
and this series in $\ell$ agrees precisely with the behavior expected from the
$\Lambda_{\rm QCD}$ ambiguity~\cite{Beneke:1998ui}. Thus Eq.~(\ref{Bup}) gives a
convergent sum-rule for the normalization $P_{1/2}$. 


$P_{1/2}$ allows us to test for a $u=1/2$ renormalon without relying on the
$n_f$-bubble chain.  Any Borel summable series of $a_n$'s in Eq.~(\ref{dm})
leads to a $P_{1/2}$ that rapidly goes to zero.  The largest physical series of
$a_n$'s that sums to $P_{1/2}=0$ has a $u=1$ renormalon, whereas $P_{1/2}\ne 0$
for any $u=1/2$ pole. Due to the universality of the $u=1/2$ renormalon of
$m_{\rm pole}$, its $P_{1/2}$ is a unique scheme independent number.  In
Fig.~\ref{sumrule} we plot the sum of terms for this $P_{1/2}$ up to $k=0$
(light/blue), $k=1$ (medium/green), and $k=2$ (dark/red).  We show the PS
(solid), static (dashed), and MSR-schemes (long-dashed), which are each
generalized to a class of schemes with $\lambda \in [1/2,2]$ using
Eq.~(\ref{lambda}).  The convergence is clearly visible, and we estimate
$P_{1/2}= 0.47\pm 0.10$. For comparison, the widely used light-fermion bubble
chain~\cite{Beneke:1998ui} (large-$n_f$ with naive-non-Abelianization, $n_f\to
-3\beta_0/2$), gives an overestimate, $P_{1/2}=0.80$.
%
%
A different series for $P_{1/2}$ was derived in
Refs.~\cite{Lee:1999ws,Pineda:2001zq}, evaluating $(1-2u)^{1+\hat b_1} B(u)$ in
an expansion about $u=0$, at $u=1/2$. It gives $P_{1/2} \simeq 0.48$, in
agreement with our result.  One can use $a_n^{\rm asym}$ in Eq.~(\ref{dm}) to
define a mass-scheme and study its $R$ dependence~\cite{Pineda:2001zq}, which
however suffers from the uncertainty in $P_{1/2}$.
 
Top-quark mass measurements from jets rely on an underlying Breit-Wigner, and
should be considered as values $m(R_0)$ in some scheme with $R_0\sim
\Gamma_t$~\cite{Fleming:2007qr}. The top $\overline {\rm MS}$ scheme has
$R\simeq 163\,{\rm GeV}\gg R_0$ so a fixed order conversion to $\overline {\rm
  MS}$ involves large logs. If we measure the MSR mass at $R_0$ and run to
$R=[\overline m(\overline m)]^{\overline {\rm MS}}$ then we directly get this
$\overline {\rm MS}$ mass.  In Fig.~\ref{rgefixed} we compare conversions
between MSR-schemes with $R_0=3\,{\rm GeV}$ and $R=163\,{\rm GeV}$, using a
fixed order expansion in $\alpha_s(\mu)=\alpha_s(\kappa R)$ (dashed curves), and
the solution of the RGE in Eq.~(\ref{NkLLsoln}) for $\gamma_i^R$ obtained with
$\mu=\kappa R$ (solid curves). Varying $\kappa$ gives a measure for the residual
uncertainty at a given order. The plot shows that Eq.(\ref{NkLLsoln}) converges
rapidly, with flat $\kappa$ dependence at NNLL. Also the RGE results display
better convergence than the standard fixed order expressions.  Comparisons using
the PS and static schemes yield the same conclusion, with similar convergence.
Taking the Tevatron mass $172.6 \pm 1.4\,{\rm GeV}$~\cite{Group:2008nq} as
$m^{\rm MSR}(R_0)$, we obtain $\overline m(\overline m)=163.0\pm 1.3
{}^{+.6}_{-.3} \pm .05 \,{\rm GeV}$. The first error is experimental, the second
takes $R_0= 3^{+6}_{-2}\,{\rm GeV}$ to account for the scheme uncertainty, and the
third is the uncertainty in our NNLL conversion. This assumes the experimental
error accounts for hadronic uncertainties.


The infrared RG analysis performed here can be generalized to study higher order
renormalons and quantities other than quark masses. For an infrared sensitive
matrix element of ${\cal O}(\Lambda_{\rm QCD}^N)$ the anomalous dimension will
have terms $R^N \alpha_s^n(R)$, and the corresponding sum-rule will provide
info on the Borel singularity at $u=N/2$.


This work was supported in part by the Department of Energy Office of Nuclear
Science under the grant DE-FG02-94ER40818, and the EU network contract
MRTN-CT-2006-035482 (FLAVIAnet). I.S. was also supported by the DOE 
OJI program and Sloan Foundation. 

\end{document}